\documentclass{appolb}
\usepackage{graphicx}
\usepackage[T1]{fontenc}
\usepackage[utf8]{inputenc}
\usepackage[polish, english]{babel}
\usepackage{caption}
\usepackage{wrapfig}
\usepackage{lipsum}
\usepackage{tikz}

\begin{document}
\title{Active target TPC for study of photonuclear reactions at astrophysical energies%
\thanks{Presented at Zakopane Conference on Nuclear Physics 2022}%
}
%
\author{M.~Kuich$^1$\thanks{e-mail: mkuich@fuw.edu.pl}, M.~Ćwiok$^1$, W.~Dominik$^1$, A.~Fijałkowska$^1$, M.~Fila$^1$, A.~Giska$^1$, Z.~Janas$^1$, A.~Kalinowski$^1$, K.~Kierzkowski$^1$, C.~Mazzocchi$^1$, W.~Okliński$^1$, M.~Zaremba$^1$, D.~Grządziel$^2$, J.~Lekki$^2$, W.~Królas$^2$, A.~Kulińska$^2$, A.~Kurowski$^2$, W.~Janik$^2$, T.~Pieprzyca$^2$, Z.~Szklarz$^2$, M.~Scholz$^2$, M.~Turzański$^2$, U.~Wiącek$^2$, U.~Woźnicka$^2$, A.~Caciolli$^3$, M.~Campostrini$^4$, V.~Rigato$^4$, M.~Gai$^5$, H.~O.~U.~Fynbo$^6$
\address{$^1$Faculty of Physics, University of Warsaw, Warsaw, Poland\\
$^2$Institute of Nuclear Physics Polish Academy of Sciences, Cracow, Poland\\
$^3$University of Padova and INFN-PD, Padova, Italy\\
$^4$Laboratori Nazionali di Legnaro, Legnaro, Italy\\$^5$University of Connecticut, CT, USA \\ $^6$Department of Physics and Astronomy, Aarhus University, Aarhus, Denmark}
}
\maketitle
\begin{abstract}
A setup designed to study photonuclear reactions at astrophysical energies - an active target Time Projection Chamber was developed and constructed at the Faculty of Physics, University of Warsaw. The device was successfully employed in two experiments at the Institute of Nuclear Physics Polish Academy of Sciences in Cracow, in which $\gamma$- and neutron-induced reactions with CO$_2$ gas target were measured. The reaction products were detected and their momenta reconstructed. Preliminary results are shown.
\end{abstract}
  
\section{Introduction}
One of the most important open questions in nuclear astrophysics concerns the creation of carbon and oxygen in the stars.
Synthesis of carbon and oxygen happens in helium-burning thermonuclear reactions in the star’s core, the triple-alpha reaction and $^{12}$C($\alpha$,$\gamma$)$^{16}$O, respectively. 
The basic observable to resolve the reaction rates is the cross-section, which has to be determined at the relevant energies (at the Gamow peak) and also is required as input information for star evolution models \cite{Fowler:1984zz}.
One way to determine the reaction cross-section is to study its inverse reaction by using an active-target together with photon and neutron beams. In this paper, we present preliminary studies to validate this approach.

\section{Experiments}
%
Studying the $\gamma$- and n-induced reactions on $^{12}$C and $^{16}$O requires a detection system that allows for reconstructing the momenta of all the charged reaction products. An ideal tool is an active-target time-projection chamber (active-target TPC), filled with CO$_2$ gas. 
The Warsaw active-target TPC was developed for this purpose. It consists of a $33\times 20 \times 20$ cm$^3$ active volume, immersed in a vacuum vessel equipped with a control system to maintain a constant gas pressure inside. The active volume is surrounded by field-shaping electrodes and terminated with a cathode plate on one end and a stack of 3 Gas Electron Multiplier foils as amplification section followed by a planar, 3-coordinate (U, V, W), redundant readout plane at the other end. The arrays of U-, V-, and W-strips register the charge deposit in 2-dimensions, while time-distribution of the charge collected at the electrode, combined with the drift velocity of the electrons in the given gas mixture and drift field, allows for determining the third coordinate. The device allows for full and unambiguous kinematic reconstruction of multiple-particle events \cite{cwiok2,Mazzocchi:2022}. The first commissioning measurements with the Warsaw active-target TPC were conducted in 2021 at the Institute of Nuclear Physics, Polish Academy of Sciences in Cracow (IFJ). 

\begin{figure}[ht!]
\begin{minipage}{0.49\textwidth}
 \includegraphics[width=0.45\textwidth, height=0.4\textwidth]{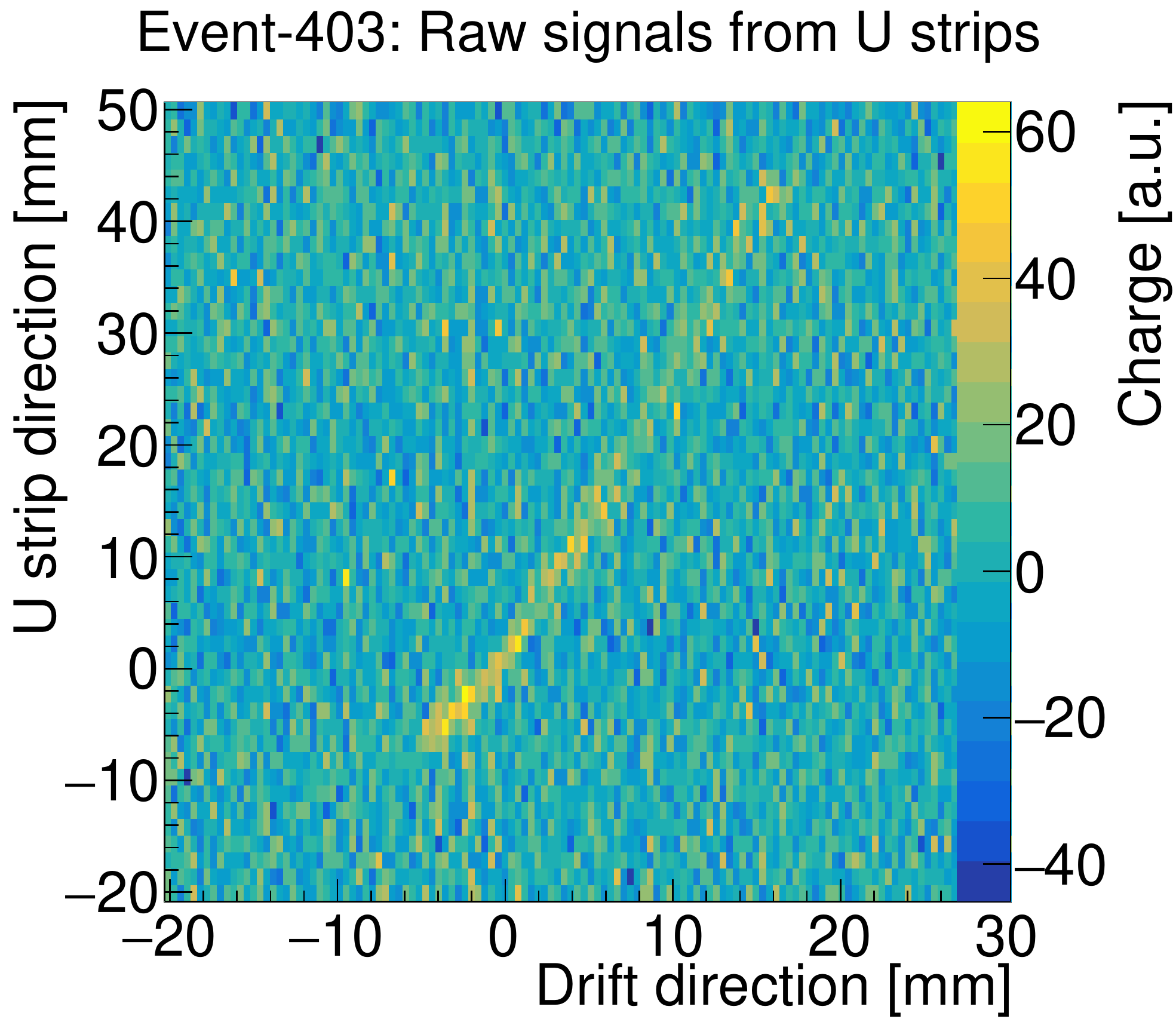}
 \includegraphics[width=0.45\textwidth, height=0.4\textwidth]{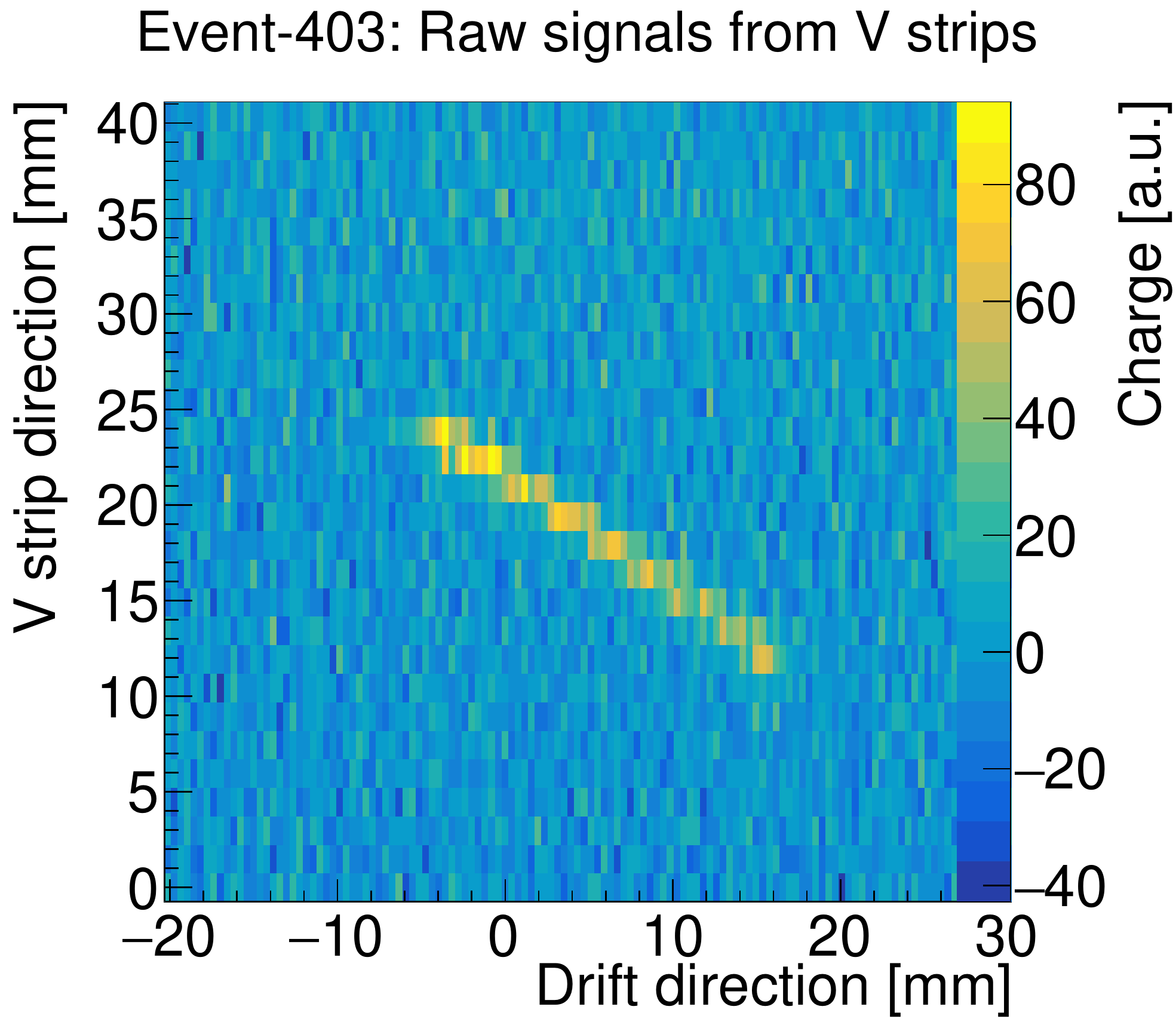}\\
  \includegraphics[width=0.45\textwidth, height=0.4\textwidth]{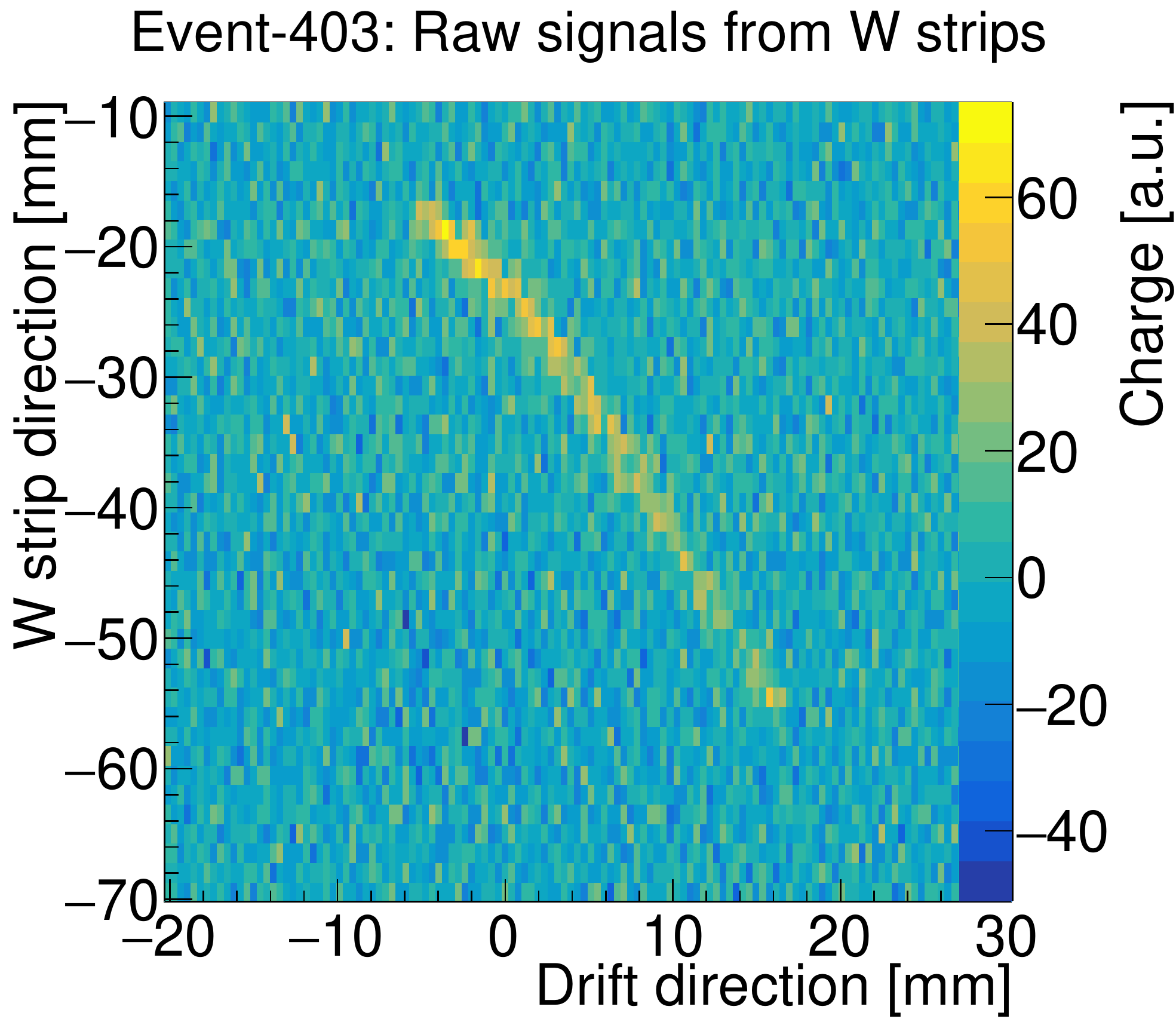}
 \includegraphics[width=0.45\textwidth, height=0.4\textwidth]{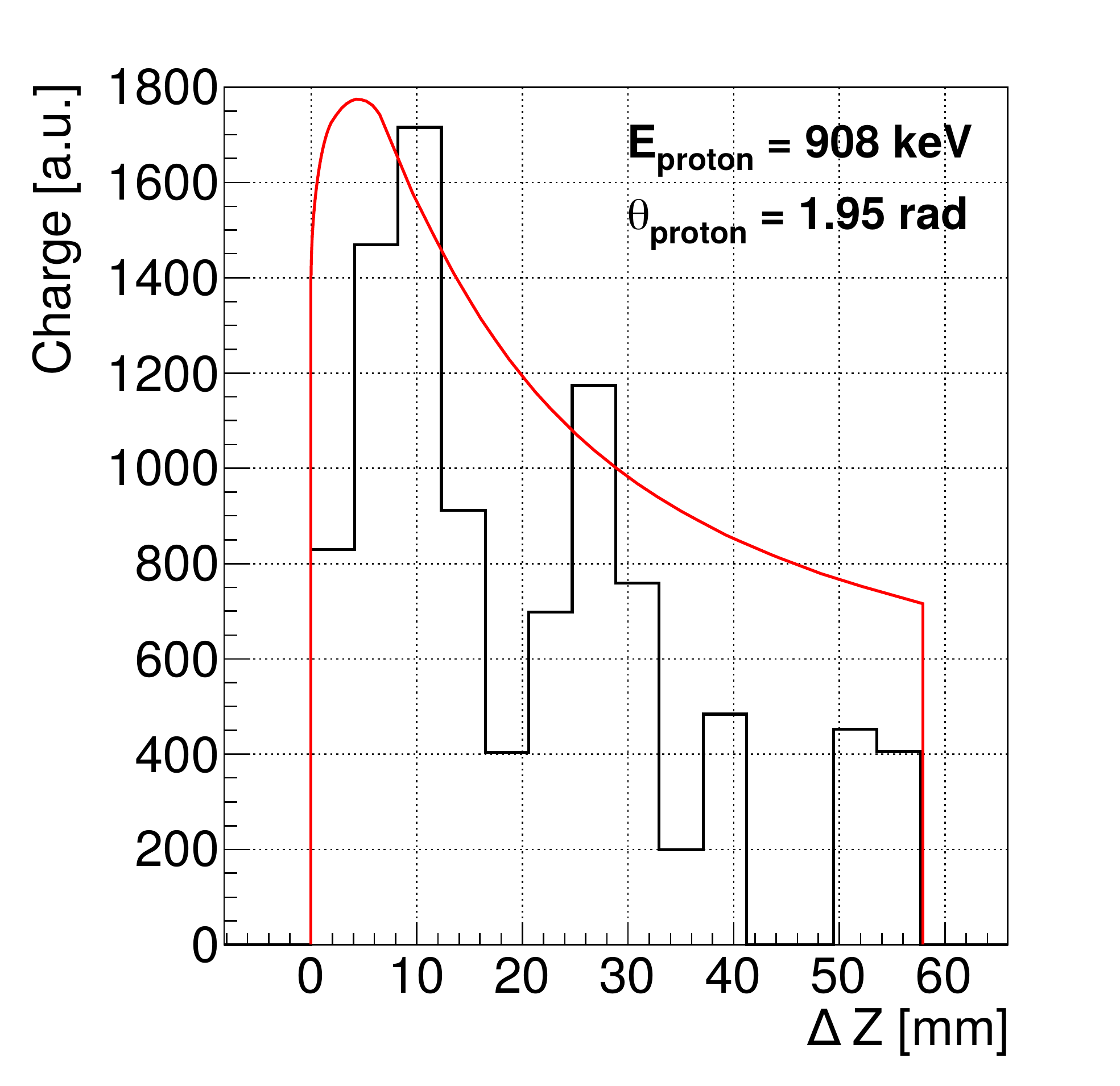}
\end{minipage}
\begin{minipage}{0.02\textwidth}
\centering
\includegraphics[width=0.3\textwidth, height=20\textwidth]{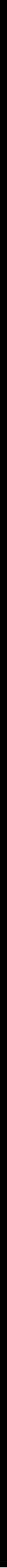}
\end{minipage}
\begin{minipage}{0.48\textwidth}
\raggedleft
 \includegraphics[width=0.45\textwidth, height=0.4\textwidth]{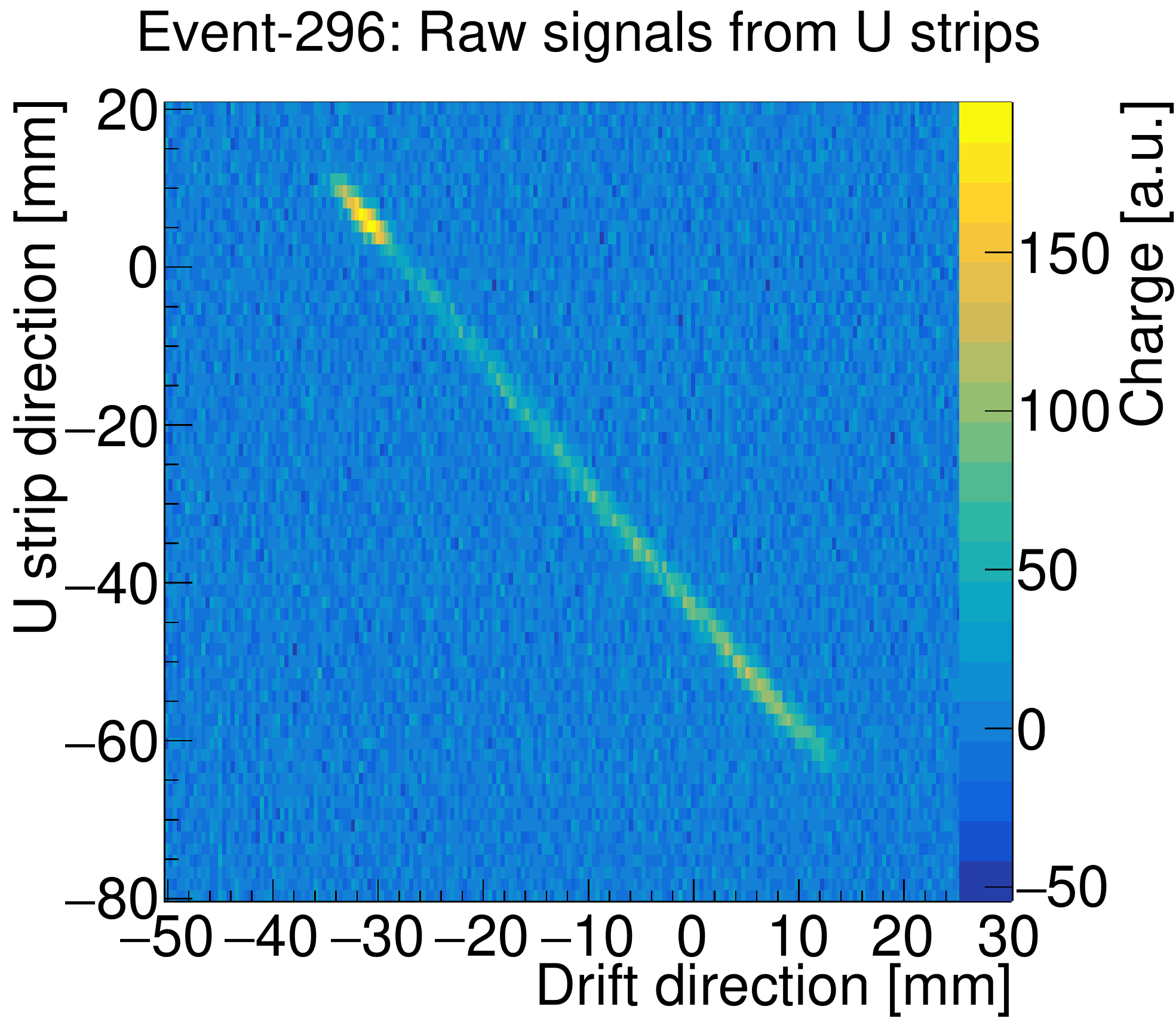}
 \includegraphics[width=0.45\textwidth, height=0.4\textwidth]{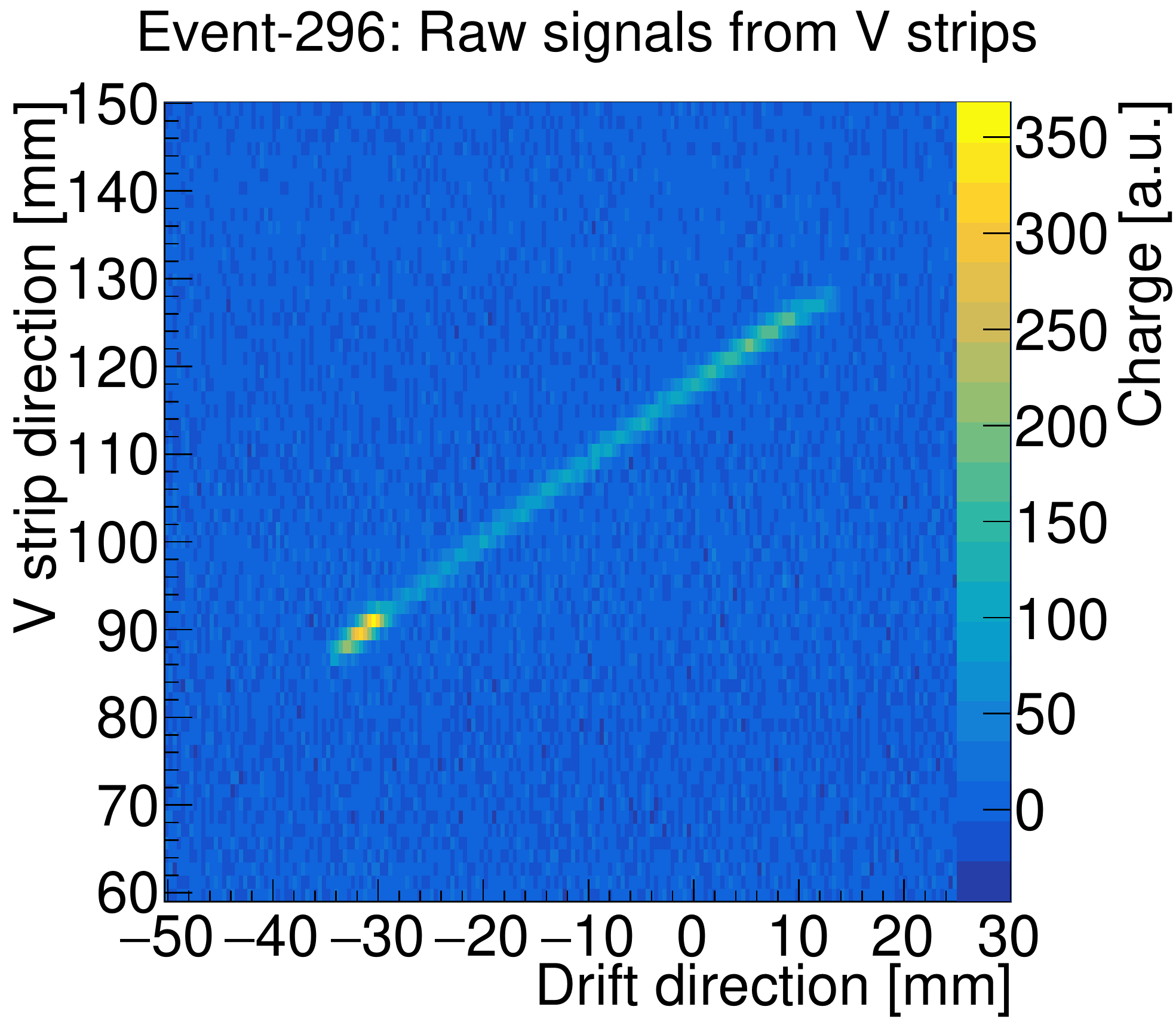}\\
  \includegraphics[width=0.45\textwidth, height=0.4\textwidth]{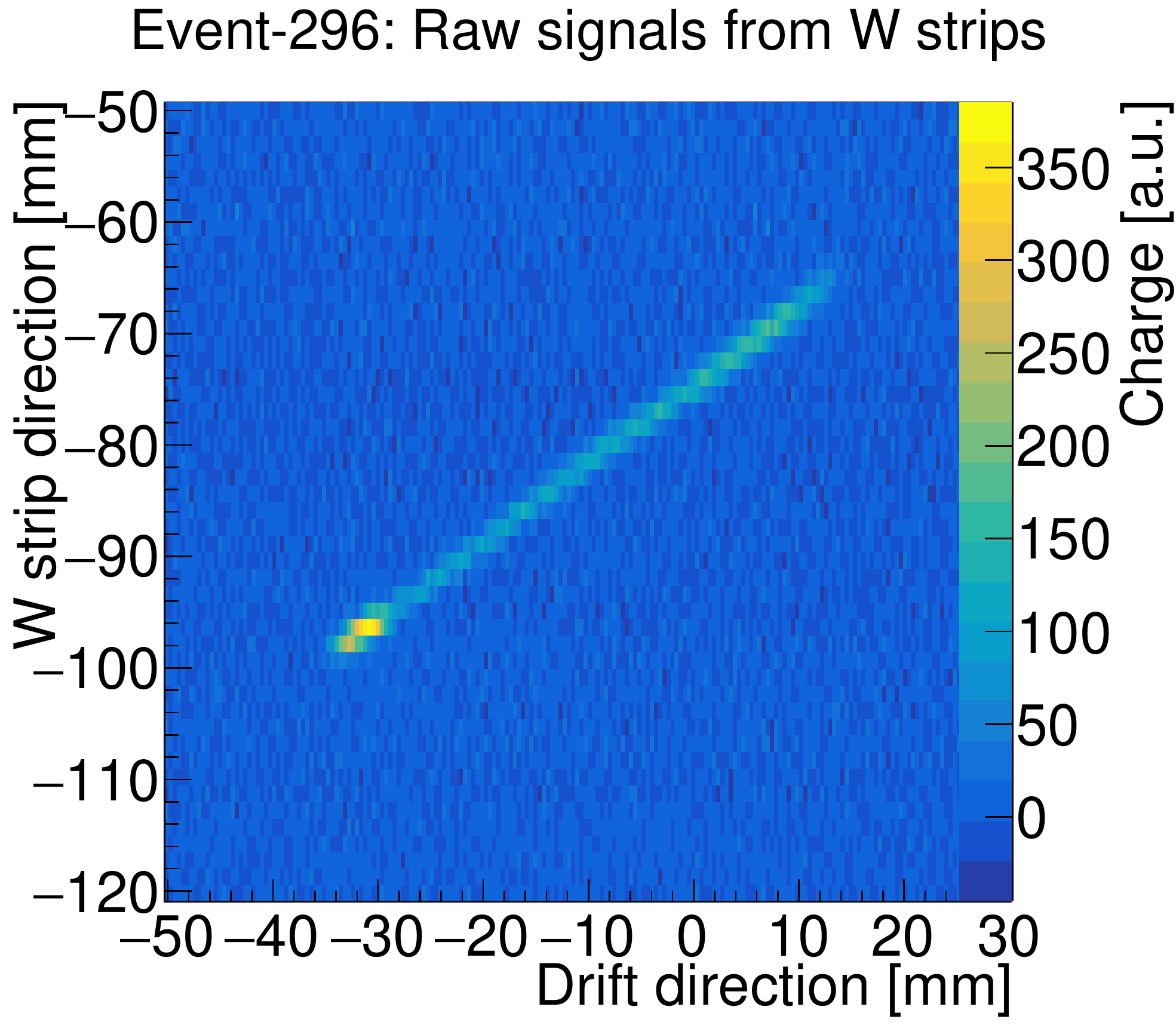}
 \includegraphics[width=0.43\textwidth, height=0.38\textwidth]{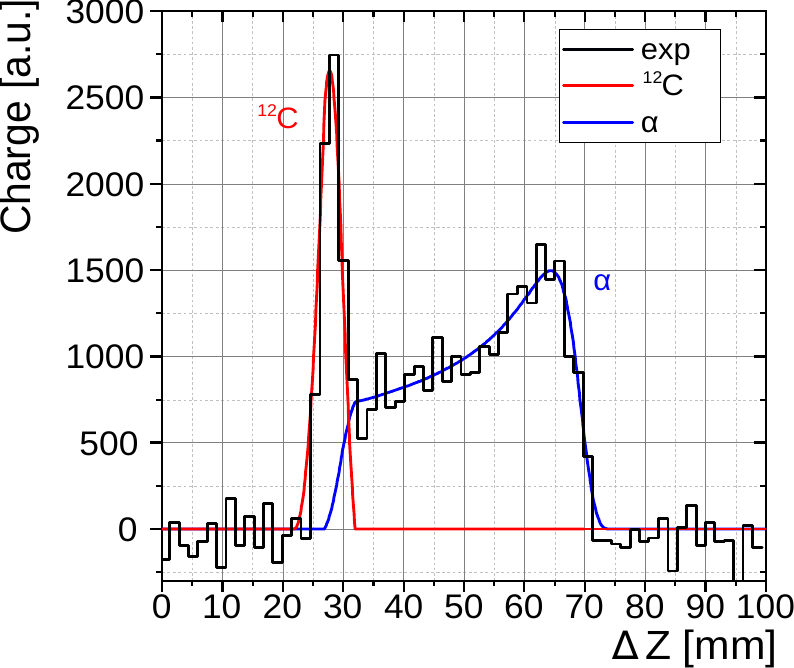}\hspace{0.1cm}
\end{minipage}\\
 \caption{Event reconstruction example for an $^{16}$O($\gamma$,p)$^{15}$N ({\it left}) and $^{16}$O($\gamma$,$\alpha$)$^{12}$C ({\it right}). The 2D plots show the raw data (U, V, W strip position vs time bin) re-scaled to mm. The bottom-right plot shows the charge distribution along the track(s) and the charge profile(s) fit.}
 \label{vdg_rec}
\end{figure}

\begin{wrapfigure}[13]{l}{0.5\textwidth}
 \includegraphics[width=0.5\textwidth, trim={0cm 0.5cm 0cm 1.3cm}, clip]{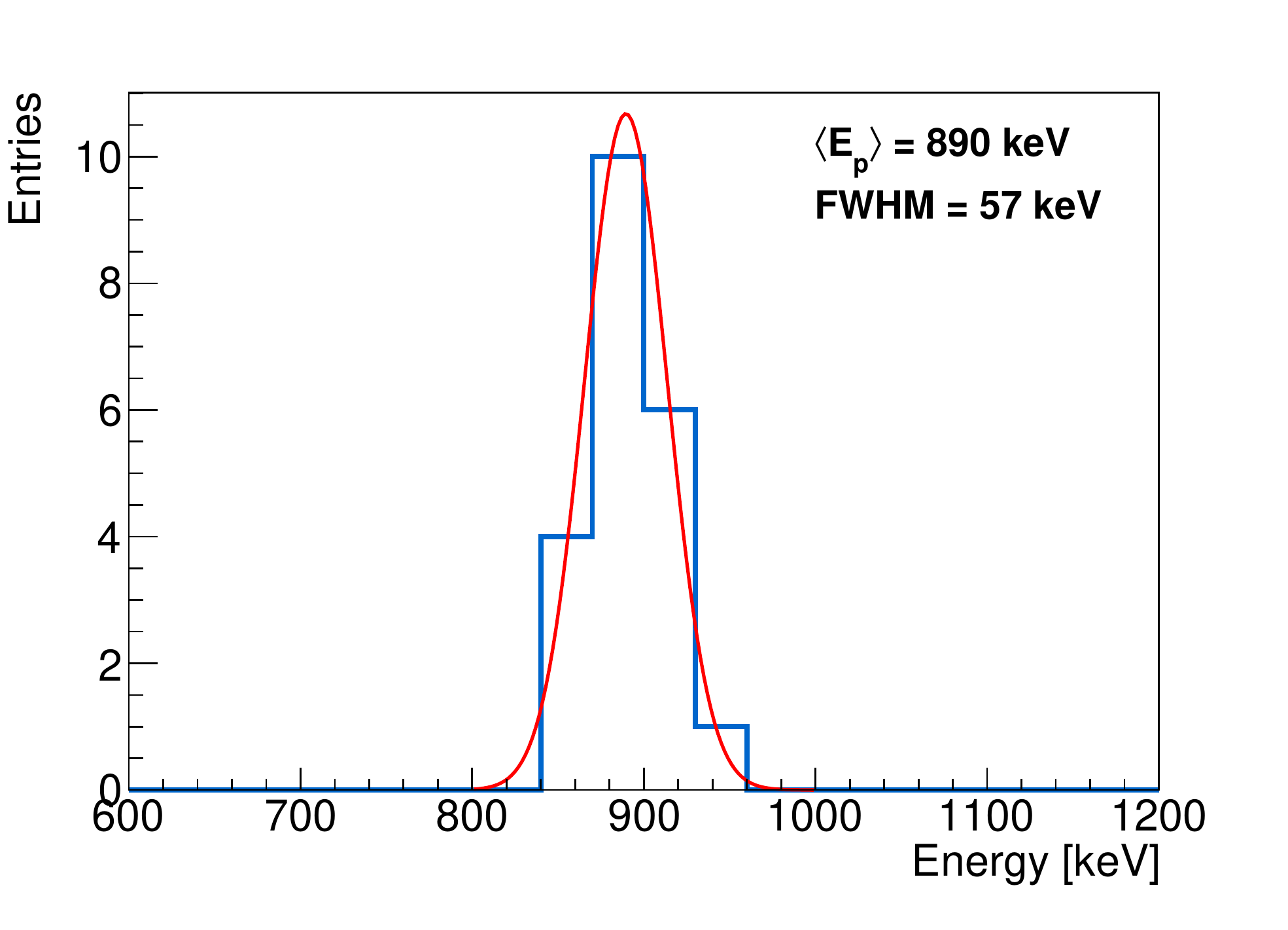}
 \caption{Reconstructed proton energy spectrum from $^{16}$O($\gamma$,p)$^{15}$N. 
 }
 \label{both}
\end{wrapfigure}

In the first experiment, a 1.03~MeV proton beam from the Van de Graaff accelerator with currents of about 10-20 $\mu$A was used to produce 13.1~MeV $\gamma$-rays in the $^{15}$N(p,$\gamma$)$^{16}$O reaction. For this purpose, a $^{15}$NCr target (about 1.3$\times 10^{18}$~atoms/cm$^2$), produced in reactive ion sputtering on a Ta backing at the National Laboratories of Legnaro, was used. The $\gamma$ beam intensity was monitored by a NaI detector positioned at the side of the target. The TPC, filled with CO$_2$ at the absolute pressure of 250 mbar, was placed right behind the target station and the produced $\gamma$-rays interacted with the gas of the TPC, where they induced the photo-disintegration of $^{12}$C and $^{16}$O. Charged reaction products were detected. Examples of $^{16}$O($\gamma$,p)$^{15}$N and $^{16}$O($\gamma$,$\alpha$)$^{12}$C event candidates are shown in Fig. \ref{vdg_rec}.  
Raw charge distributions in strip (U-, V-, W-) coordinate and time are presented as 2D plots. A manual reconstruction method was used to determine the emitted particle position in the chamber and measure its range. Identification of the two-particle event as due to $^{16}$O($\gamma$,p)$^{15}$N or $^{16}$O($\gamma$,$\alpha$)$^{12}$C as well as the energy reconstruction was done by comparing the track length of emitted particles with SRIM simulations \cite{srim}. Examples of event reconstruction is presented in the 1D plots in Fig.~\ref{vdg_rec} as charge distribution along the track(s) with fitted Bragg curve(s). 
A preliminary distribution of the reconstructed energy of the protons is depicted in Fig.\ref{both} (partial statistics was analyzed).
The mean energy of the protons, $\langle E_\textup{p} \rangle = 890 \pm 24$~keV, is in 3$\sigma$ agreement with expected 966~keV.

\begin{figure}[ht!]
\begin{minipage}{0.49\textwidth}
 \includegraphics[width=0.45\textwidth, height=0.4\textwidth]{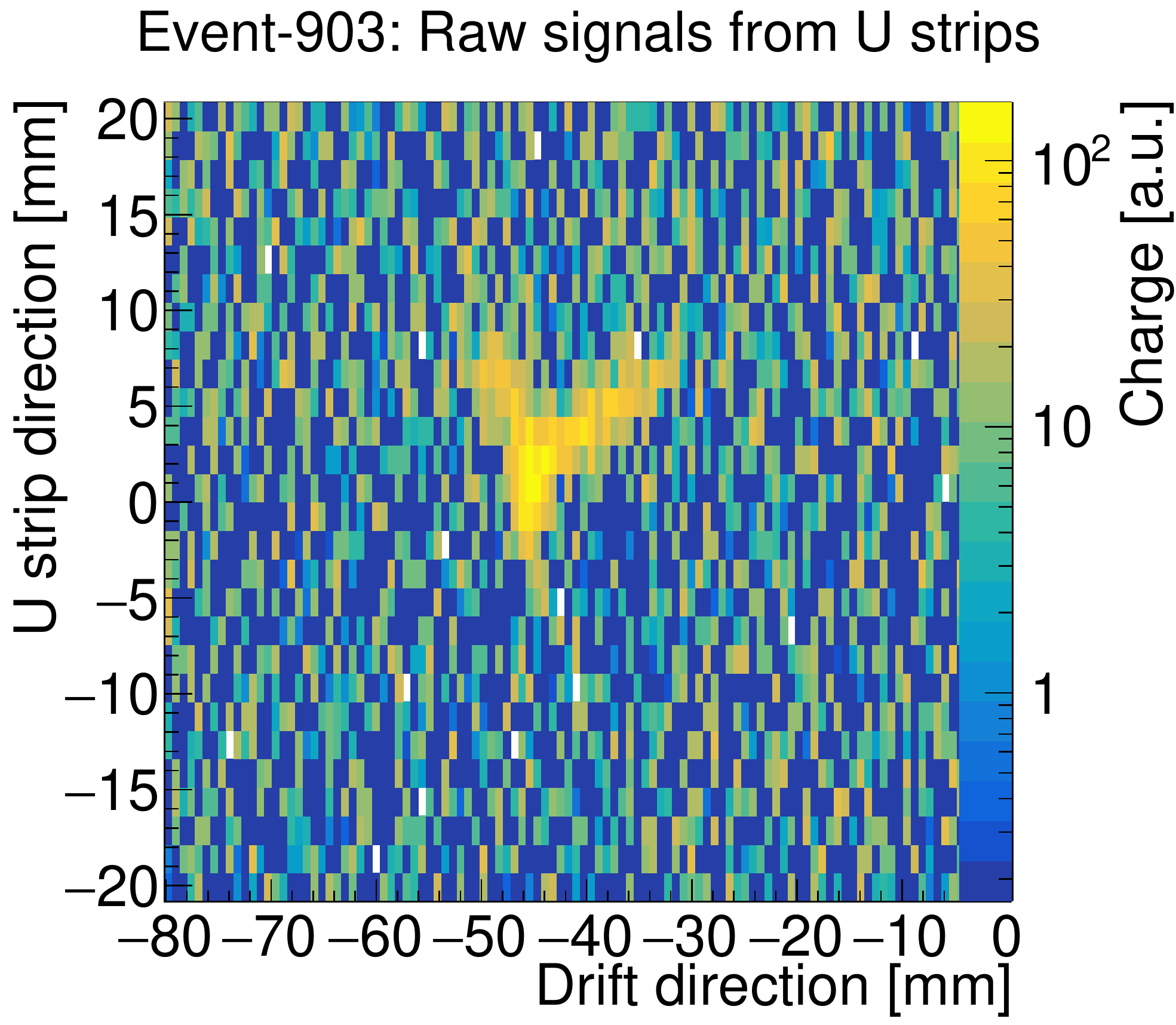}
 \includegraphics[width=0.45\textwidth, height=0.4\textwidth]{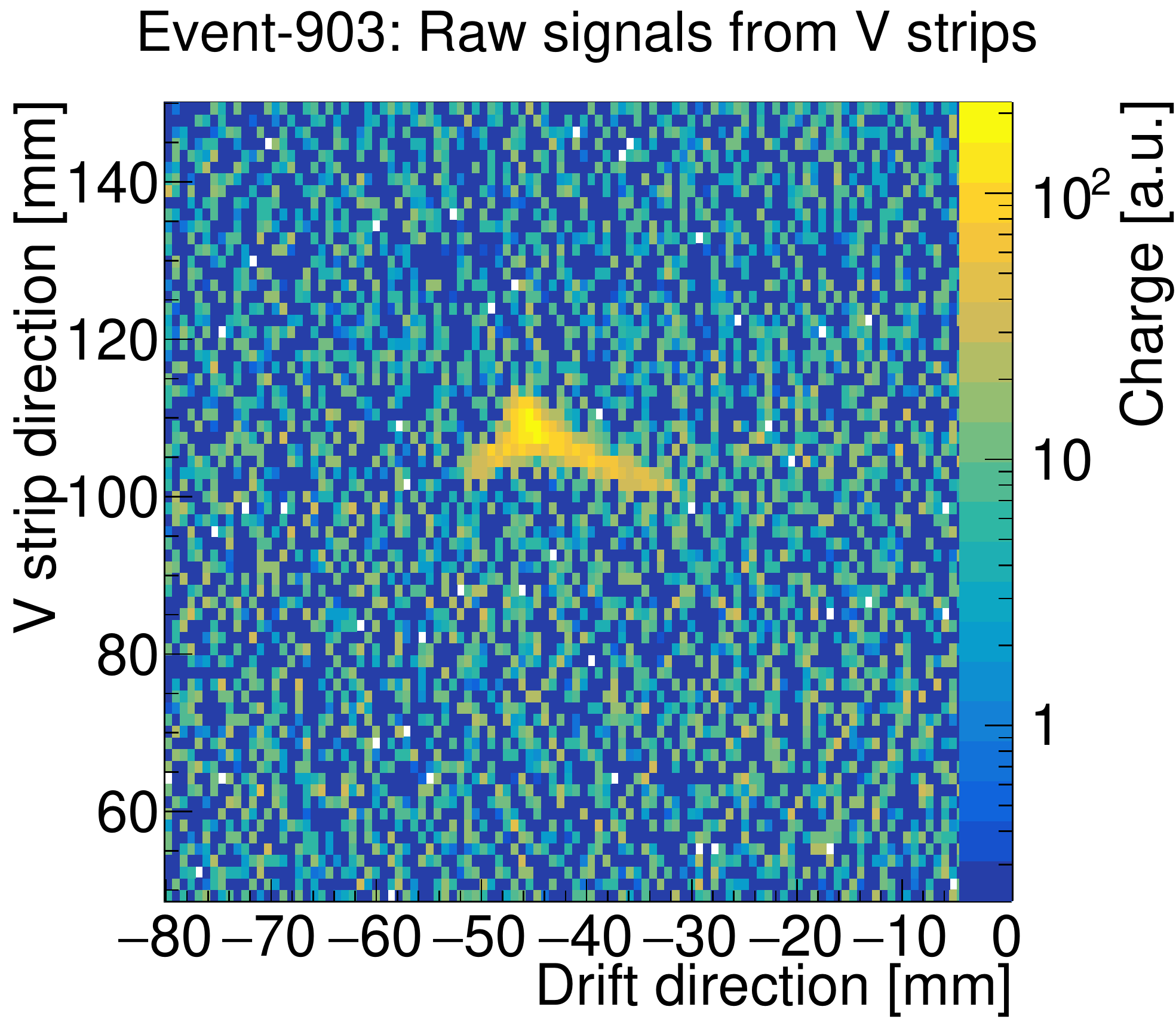}\\
  \includegraphics[width=0.45\textwidth, height=0.4\textwidth]{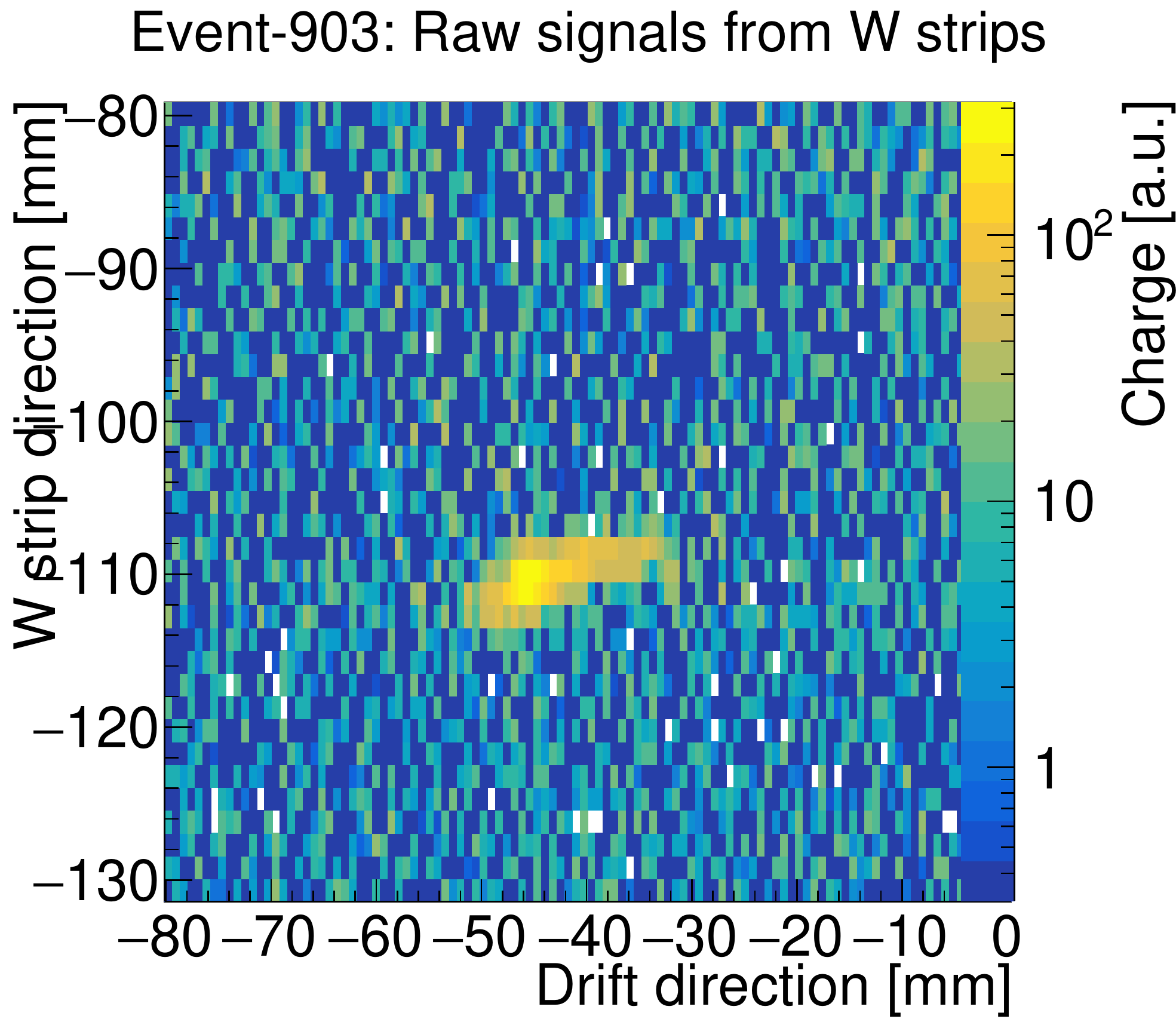}
 \includegraphics[width=0.43\textwidth, height=0.38\textwidth]{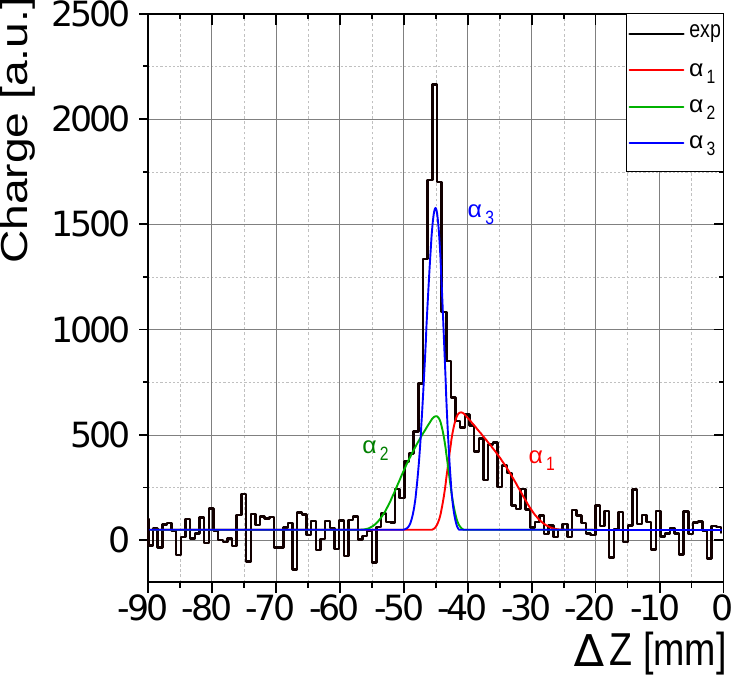}
\end{minipage}
\begin{minipage}{0.02\textwidth}
\centering
\includegraphics[width=0.3\textwidth, height=20\textwidth]{figures/line.pdf}
\end{minipage}
\begin{minipage}{0.49\textwidth}
\raggedleft
 \includegraphics[width=0.45\textwidth, height=0.4\textwidth]{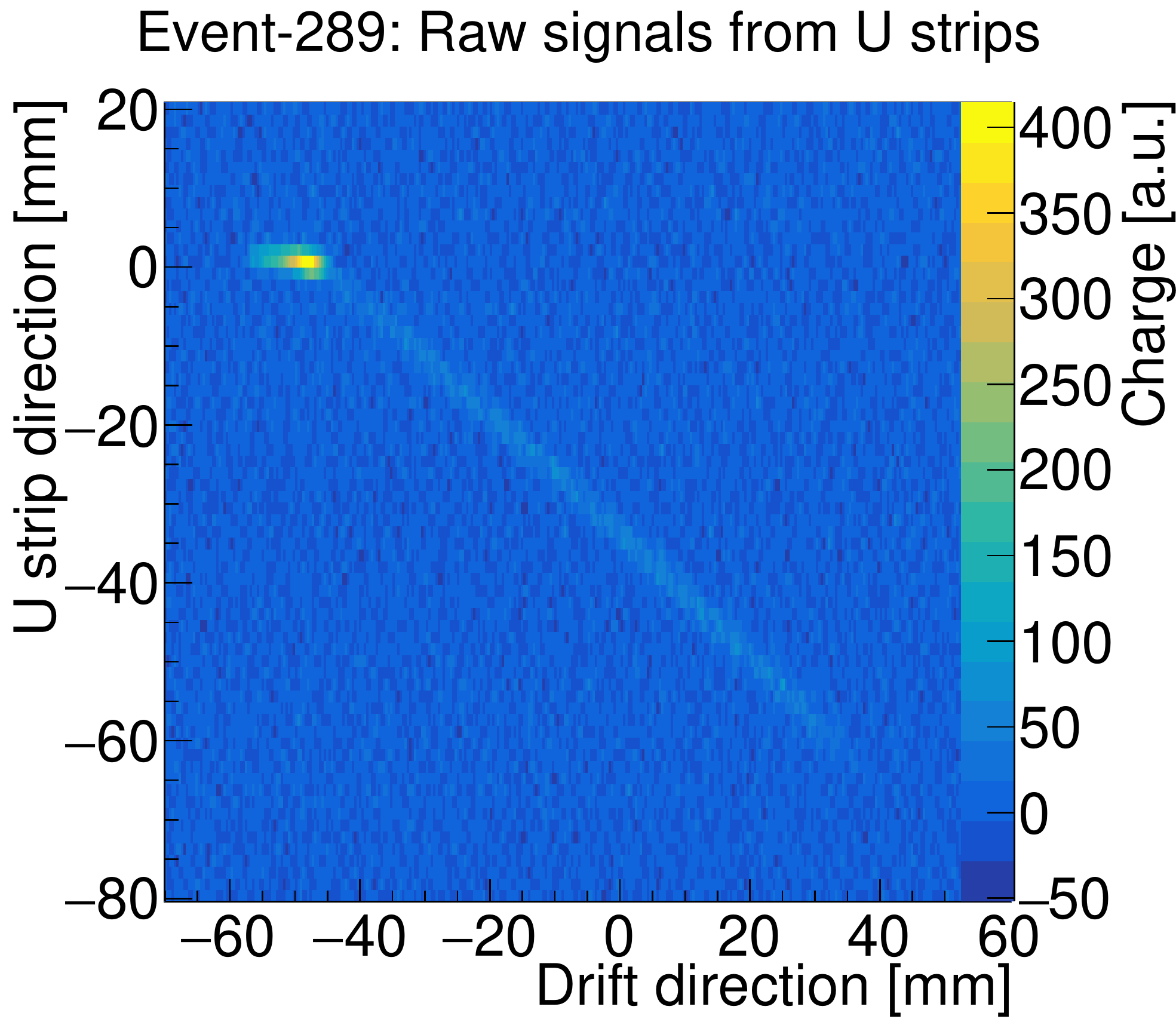}
 \includegraphics[width=0.45\textwidth, height=0.4\textwidth]{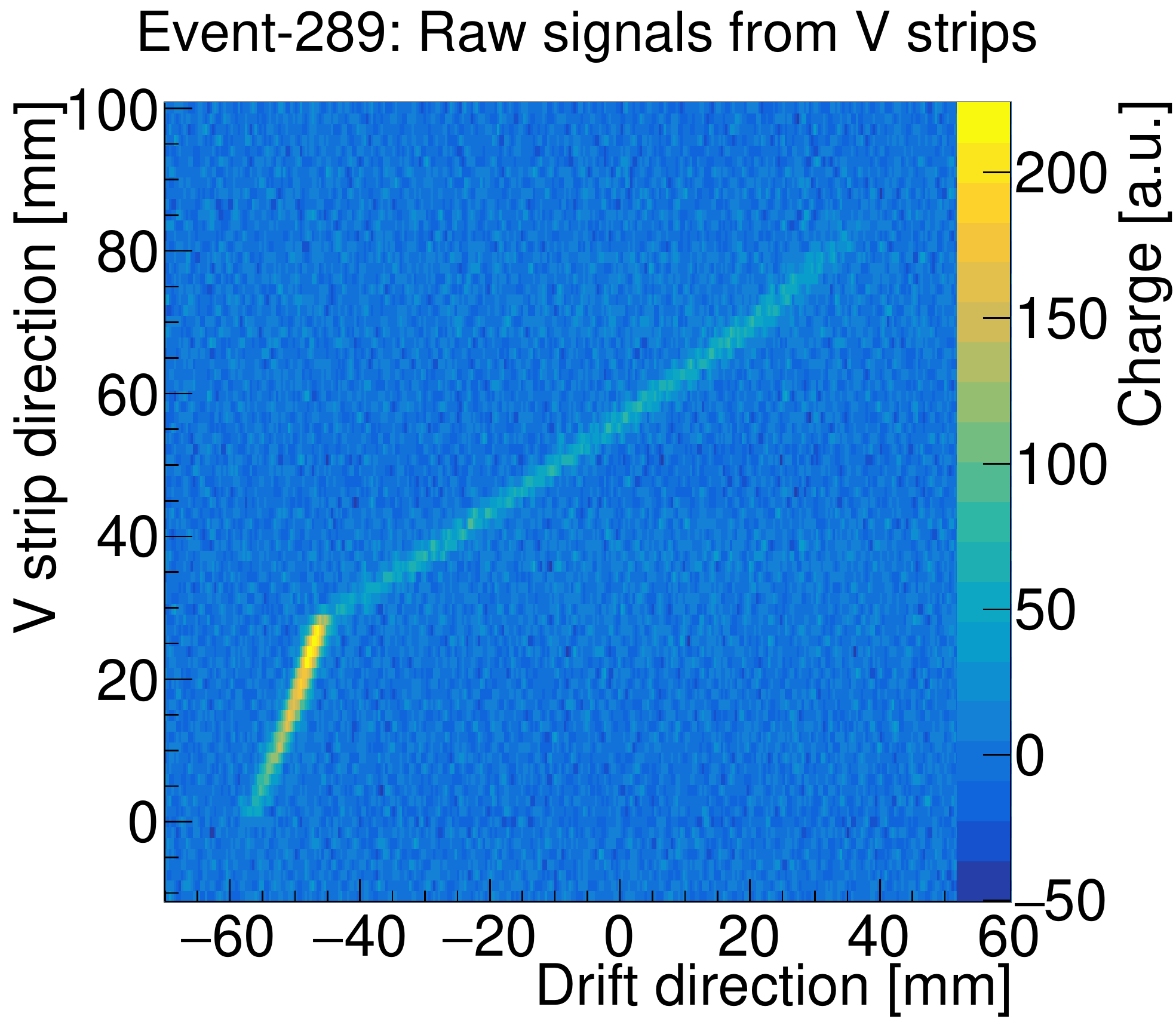}\\
  \includegraphics[width=0.45\textwidth, height=0.4\textwidth]{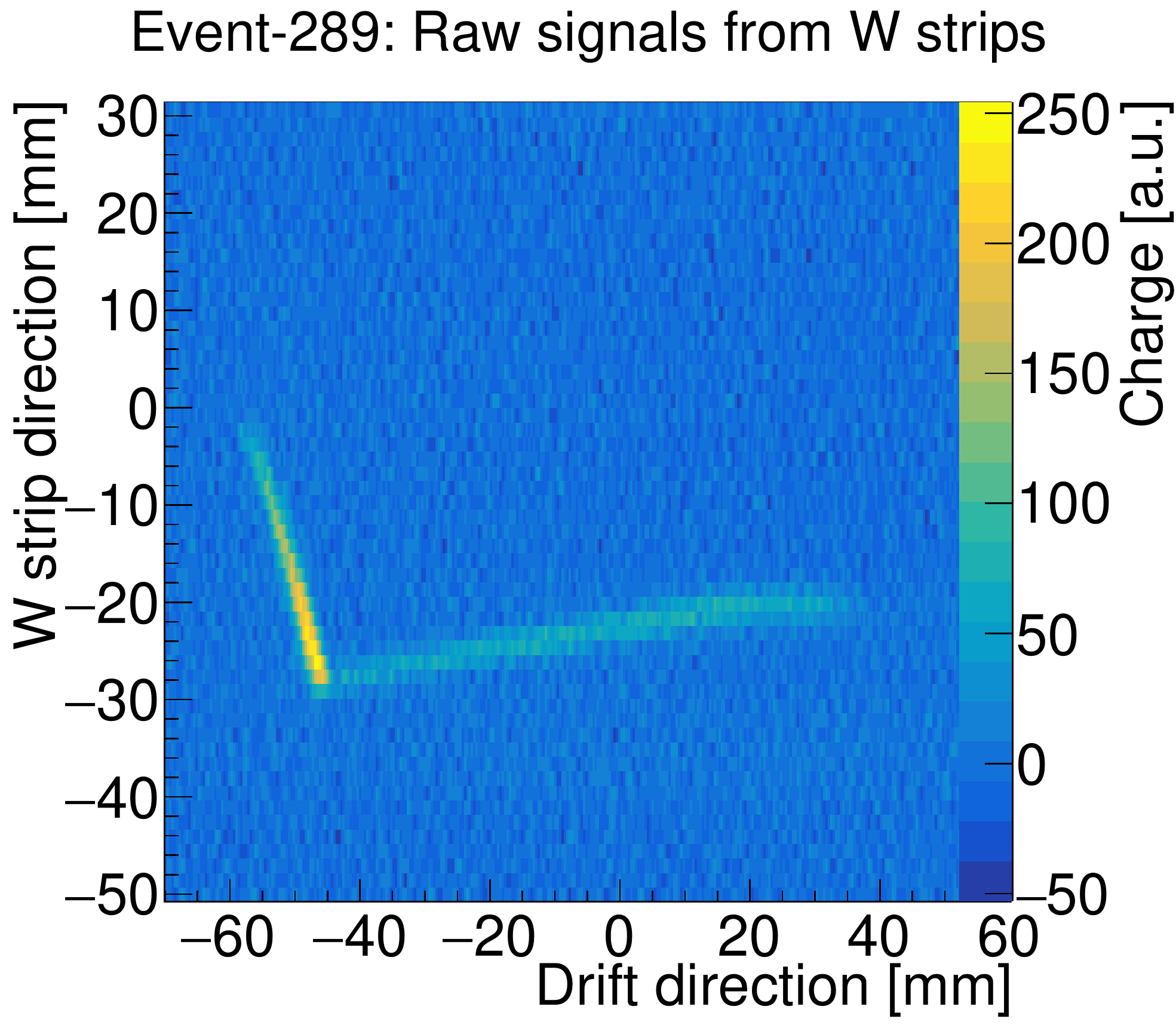}
 \includegraphics[width=0.43\textwidth, height=0.38\textwidth]{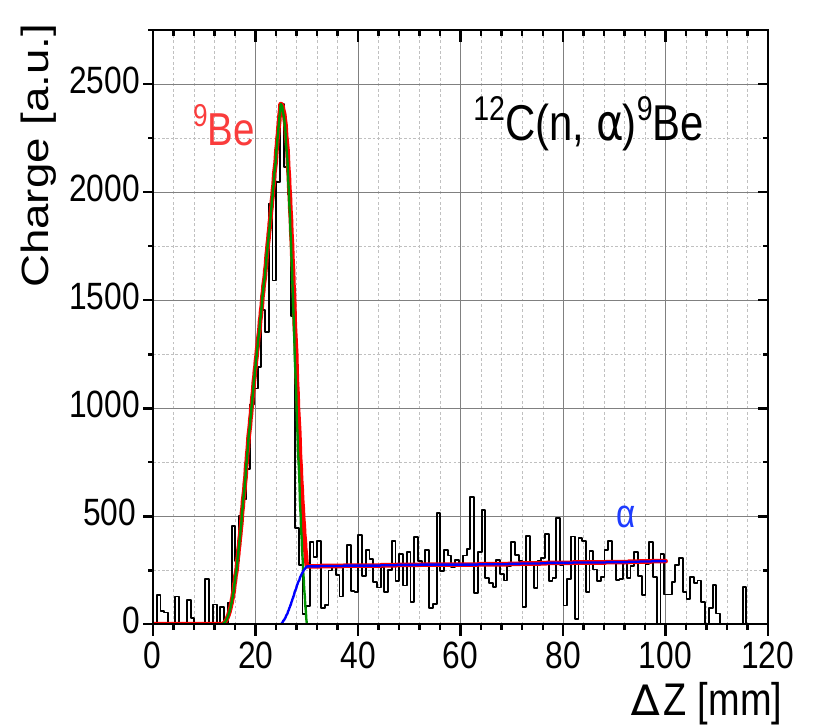}\hspace{0.1cm}
\end{minipage}\\
 \caption{Event reconstruction example for an $^{12}$C(n, n')$^{12}$C$^*$ ({\it left}) and $^{12}$C(n,$\alpha$)$^{9}$Be ({\it right}). The 2D plots show the raw data (U, V, W strip position vs time bin) re-scaled to mm. The bottom-right plot shows the charge distribution along the track(s) and the charge profile(s) fit.}
 \label{ign_rec}
\end{figure}

The second experiment took place at the Impulse Neutron Generator (IGN-14), where 14~MeV neutrons were produced in the d(t,n) reaction and interacted with the CO$_2$ gas, which was kept at 80~mbar, in the TPC positioned just behind the tritium target. The neutron flux was monitored with a $^3$He counter and additionally estimated with activation of aluminium targets.
The typical neutron flux amounted to about $3 \times 10^4$~n/s/cm$^2$.
Two example events for $^{12}$C(n,n')$^{12}$C$^*$ and $^{16}$O(n,$\alpha$)$^{9}$Be reactions are presented in Fig.~\ref{ign_rec}.
Partial statistics was manually analyzed and the emitted particles identified. The majority of the two-particle events were classified as $^{12}$C(n,$\alpha$)$^{9}$Be, while three-particle events as carbon dissociation into 3 $\alpha$ particles. Preliminary reconstruction of the particles momenta indicates that the $^{12}$C(n,n')$^{12}$C$^*$ event in the Fig.~\ref{ign_rec} corresponds to the decay of the Hoyle state at 7654 keV by 3$\alpha$ emission.

\section{Summary}
An active-target TPC dedicated for studying reactions of astrophysical interest at the relevant energies with $\gamma$ or neutron beams was developed at the University of Warsaw. The first experiments employing it were conducted in 2021 at IFJ, where $^{16}$O($\gamma$,$\alpha$)$^{12}$C, $^{16}$O($\gamma$,p)$^{15}$N, $^{12}$C($\gamma$,$3\alpha$), $^{12}$C(n,$\alpha$)$^9$Be and $^{12}$C(n,n')$^{12}$C$^*$ were observed. Preliminary analysis shows that the Warsaw active-target TPC is an adequate tool for measuring such reactions induced by non-charged particles.

\bigskip
\noindent
We would like to thank H. Czyrkowski and R. Dąbrowski for their support in the preparation of the equipment. Scientific work was supported by the National Science Centre, Poland, contract no. 2019/33/B/ST2/02176, by the University of Warsaw, Poland, through the Interdisciplinary Centre for Mathematical and Computational Modelling, comp. alloc. no. G89-1286. 

\bibliography{bib}{}

\begin{thebibliography}{1}

\bibitem{Fowler:1984zz}
W.~A. Fowler.
\newblock {\em Rev. Mod. Phys.}, 56:149, 1984.

\bibitem{cwiok2}
M.~Ćwiok et~al.
\newblock {\em Acta Phys.Pol. B}, 49:509, 2018.

\bibitem{Mazzocchi:2022}
M.~Ćwiok et~al.
\newblock {\em submitted to EPJ Web of Conferences 2022}.

\bibitem{srim}
J.~F. Ziegler et~al.
\newblock {\em Nucl. Instr. and Meth. in Phys. Res. B}, 268(11):1818, 2010.

\end{thebibliography}
  \bibliographystyle{unsrt}

\end{document}